\documentclass[epj,twocolumn]{webofc}
\usepackage[varg]{txfonts}   
\usepackage{epsfig}
\usepackage{amssymb,amsmath}

\woctitle{Hadron Collider Physics symposium 2012}
\begin{document}
\title{Top Forward-backward asymmetry at the Tevatron vs. Charge asymmetry 
\\ at the LHC in chiral $U(1)'$ models with flavored Higgs fields}
%
%

\author{P. Ko\inst{1}\fnsep\thanks{\email{pko@kias.re.kr}} \and
        Yuji Omura\inst{2}\fnsep\thanks{\email{yujiomura3@gmail.com}} \and
        Chaehyun Yu\inst{1}\fnsep\thanks{\email{chyu@kias.re.kr}}
}

\institute{School of Physics, KIAS, Seoul 130-722, Korea
\and
     Physik Department T30, Technische Universit\"{a}t M\"{u}nchen, 
James-Franck-Stra$\beta$e, 85748 Garching, Germany  
}

\abstract{%
An extra $U(1)'$ model with $Z'$ coupled only to the right-handed (RH) up-type 
quarks has been one of the popular models for the Tevatron top forward-backward asymmetry (FBA),   and has been excluded by the same-sign top-pair productions 
at the LHC. However, the original $Z'$ model is not physical, 
since the up-type  quarks are massless including the top quark. 
This disaster can be evaded if the Higgs sector is extended by including 
new Higgs doublets with nonzero $U(1)'$ charges.   
We find that some parameter regions could achieve not only the top FBA 
at the Tevatron, but also the charge asymmetry at the LHC without exceeding 
the upper limit of the same-sign top-quark pair production at the LHC.
The lesson is that it is mandatory to extend the Higgs sector whenever one
considers chiral gauge symmetries beyond the SM gauge group. Otherwise 
some fermions remain massless, and thus it is meaningless to work on 
phenomenology without the extra Higgs doublets with new chiral 
gauge charges.
}
\maketitle
\section{Introduction}
The top forward-backward asymmetry ($A_\textrm{FB}^t$) is one of the most
interesting observables because there exists discrepancy
between theoretical predictions in the standard model (SM) 
and experimental results at the Tevatron. The most recent measurement
for $A_\textrm{FB}^t$ at CDF is 
$A_\textrm{FB}^t=0.162\pm 0.047$ in the letpon+jets channel with a full
set of data~\cite{cdfnew}, which is consistent with
the previous measurements at CDF and D0 within uncertainties~\cite{oldafb}. 
The SM predictions are between $0.06$ and $0.09$~\cite{smafbac,smafb},
so that the deviation is around $2 \sigma$.

If the discrepancy in $A_\textrm{FB}^t$ is generated by new physics,
the new physics model would be tested at the LHC. 
One of the good measurements is the charge asymmetry $A_C^y$, which is 
defined by the difference of numbers of events with the positive and negative 
$\Delta |y|=|y_t|-|y_{\bar{t}}|$ divided by their sum.
The current values for $A_C^y$ are
$A_C^y=-0.018\pm 0.028\pm 0.023$ at ATLAS~\cite{atlasacy} and 
$A_C^y=0.004\pm 0.010\pm 0.012$ at CMS~\cite{cmsacy}, respectively, 
which are consistent with the SM prediction $\sim 0.01$~\cite{smafbac}. 
Another interesting observable at the LHC is the cross section for
the same-sign top-quark pair 
production, $\sigma^{tt}$, which is not allowed in the SM. 
The current upper bound on $\sigma^{tt}$ is about 17 pb at CMS~\cite{cmssame}
2 pb or 4 pb at ATLAS depending on the model~\cite{atlassame}.
\footnote{Very recently the CMS Collaboration updated it to be less than 
0.39 pb at 95 \% C.L..~\cite{Chatrchyan:2012sa}.}
Some models which were proposed to account for $A_\textrm{FB}^t$ at the Tevatron,
predict large $A_C^y$ and/or $\sigma^{tt}$ so that they are already disfavored
by present experiments at the LHC.

\section{The original $Z^{'}$ model by Jung {\it et al.} \cite{zprime}}
Let us first consider a $Z'$ model first proposed by Jung, Murayama, Pierce and 
Wells ~\cite{zprime}.  In this model, it is assumed that there is a flavor changing
couplings of $Z'$ to the right-handed (RH) $u$ and $t$ quarks:
\begin{equation}
{\cal L} = - g_X Z_\mu^{'} \left[  \overline{t_R} \gamma^\mu u_R + H.c. \right] .
\end{equation}
The $t$-channel exchange of $Z'$ leads to the Rutherford peak in the forward direction
and generates the desired amount of the top FBA if $Z'$ is around $150-250$ GeV and
$g_X$ is not too small. Here $Z'$ is assumed to couple only to the right-handed (RH) 
quarks in order to evade the strong bounds from the FCNS processes such as 
$K^0 - \overline{K^0}$,
$B^0_{d(s)} - \overline{B_{d(s)^0}}$ mixings and $B \rightarrow X_s \gamma$. 
And such a  light $Z^{'}$ should be leptophobic in order to avoid the strong bounds 
from the Drell-Yan  processes. Therefore the original $Z'$ model is chiral, leptophobic 
and flavor non-universal.  One can imagine that $Z'$ is associated with a new local 
gauge symmetry $U(1)'$. Then the original $Z'$ model has gauge anomalies and 
is not consistent.  Also one can not write Yukawa couplings for the up-type quarks 
if we have only the SM Higgs doublet which has the vanishing $U(1)'$ charge.
Therefore it would be highly nontrivial to construct a realistic gauge theory which 
satisfies the conditions in the original $Z'$ model.  Also the original $Z'$ model was 
excluded by the same sign top pair productions, because $Z'$ exchange can contribute 
to $ u u \rightarrow t t$.  The  upper bounds on the same-sign top-pair production put
strong constraints on this model~\cite{saavedra}. 
However the story is not that simple for various reasons described above, and the 
model should be extended with new Higgs doublets as described in the next section
~\cite{u1models}.

\section{$U(1)'$ models with flavored multi-Higgs doublets by Ko, Omura and Yu
~\cite{u1models}}


In this section we review the flavor-dependent chiral U(1)$^\prime$ model 
with flavored Higgs doublets that were proposed in Ref.~\cite{u1models}. 
Our model is an extension of the $Z^\prime$ model ~\cite{zprime} described in the previous 
section, curing various problems of Ref.~\cite{zprime}.  
The $Z^\prime$ boson must be associated with some gauge symmetry 
if we work in weakly interacting theories, and we consider an extra U(1)$^\prime$ symmetry
~\cite{u1models}.  The $Z^\prime$ boson better be leptophobic to avoid
the stringent constraints from the LEP II and Drell-Yang experiments.
Furthermore, it would be very difficult to assign flavor-dependent
U(1)$^\prime$ charges to the down-type quarks and left-handed up-type quarks
because it gives rise to dangerous FCNCs.
Therefore we assigned flavor-dependent  U(1)$^\prime$ charges $u_i$ $(i=u,c,t)$
only to the right-handed up-type quarks while the left-handed quarks and
right-handed down-type quarks are not charged under $U(1)'$.  

Then, the Lagrangian between $Z^\prime$ and the SM quarks 
in the interaction eigenstates is given by
\begin{equation}
\mathcal{L}_{Z^\prime q\bar{q}} =
g^\prime \sum_i u_i  Z^\prime_\mu \overline{U_R^i} \gamma^\mu U_R^i,
\end{equation}
where $U_R^i$ is a right-handed up-type quark field in the interaction eigenstates 
and $g^\prime$ is the couping of the U(1)$^\prime$. 

After the electroweak symmetry breaking, we can rotate the quark fields 
into the mass eigenstates by bi-unitary transformation.   
The interaction Lagrangian for the $Z^\prime$ boson in the mass eigenstate
is given by
\begin{eqnarray}
\mathcal{L}_{Z^\prime q\bar{q}}  & =  & 
g^\prime Z^\prime_\mu \left[  
(g_R^u)_{ut} \overline{u_R} \gamma^\mu t_R
+(g_R^u)_{ut} \overline{t_R} \gamma^\mu u_R
\right.   \nonumber  \\
& & +  \left.(g_R^u)_{uu} \overline{u_R} \gamma^\mu u_R
+(g_R^u)_{tt} \overline{t_R} \gamma^\mu t_R
\right].
\end{eqnarray}
The $3\times 3$ mixing matrix $(g_R^u)_{ij} = (R_u)_{ik} u_k (R_u)^\dagger_{kj}$ 
is the product of the U(1)$^\prime$ charge matrix ${\rm diag} ( u_{k=1,2,3} )$ and
a unitary matrix $R_u$, where the matrix  $R_u$ relates the RH up-type quarks 
in the interaction eigenstates and in the mass eigenstates.   
The matrix $R_u$ participates in diagonalizing the up-type quark mass matrix. 
We note that the components of the mixing angles related 
to the charm quark have to be small in order to respect 
constraints from the $D^0$-$\overline{D^0}$ mixing.

If one assigns the U(1)$^\prime$ charge $(u_i)=(0,0,1)$
to the right-handed up-type quarks, one can find the relation 
$(g_R^u)_{ut}^2 = (g_R^u)_{uu} (g_R^u)_{tt}$ \footnote{We note 
that the relation is not valid for the other charge assignments. For general cases, 
we introduce a parameter $\xi$ with $(g_R^u)_{uu} (g_R^u)_{tt} = \xi (g_R^u)_{ut}^2 $
where $\xi$ is a free parameter. }. 
This relation indicates that if the $t$-channel diagram mediated by $Z^\prime$ 
contributes to the $u\bar{u}\to t\bar{t}$ process, the $s$-channel diagram 
mediated by $Z^\prime$ should be taken into account, too.

As we discussed in the previous section, it is mandatory to include
additional flavored Higgs doublets charged under U(1)$^\prime$ in order 
to write down proper Yukawa interactions for the SM quarks charged
under U(1)$^\prime$ at the renormalizable level \footnote{It is also true that 
one cannot write nonrenormalizable Yukawa interactions with  the SM 
Higgs doublet only. It is essential to include the Higgs doublets with nonzero 
$U(1)^{'}$ charges in order that one can write Yukawa couplings for the up-type
quarks in this model.}.
The number of additional Higgs doublets depends on the U(1)$^\prime$ charge 
assignment to the SM fermions,  especially the right-handed up-type quarks. 
In general, one must add three additional Higgs doublets with U(1)$^\prime$ 
charges $u_i$ (see Ref.~\cite{u1models} for more discussions).
For the charge assignment $(u_i)=(0,0,1)$ we have two Higgs doublets
including the SM-like Higgs doublet, while for $(u_i)=(-1,0,1)$ three Higgs
doublets are required. The additional U(1)$^\prime$ must be broken in the end, 
so that we add a U(1)$^\prime$-charged singlet Higgs field $\Phi$ to the SM. 
Both the U(1)$^\prime$-charged Higgs doublet and the singlet $\Phi$ can give
the masses for the $Z^\prime$ boson and extra fermions if it has
a nonzero vacuum expectation value (VEV). After breaking of the electroweak 
and U(1)$^\prime$ symmetries, one can write down the Yukawa interactions 
in the mass basis.
After all the Yukawa couplings would be proportional to the quark 
masses responsible for the interactions so that we could ignore 
the Yukawa couplings which are not related to the top quark. 

The number of relevant Higgs bosons participating in the top-quark pair 
production depends on the U(1)$^\prime$ charge assignment and mixing angles. 
The relevant Yukawa couplings for the top-quark pair production can be written as
\begin{equation}
V =  Y_{tu}^h \overline{u_L} t_R h+Y^H_{tu} \overline{u_L} t_R H 
+ i Y_{tu}^a \overline{u_L} t_R a + h.c.,
\end{equation}
where $h$ and $a$ are the lightest neutral scalar and pseudoscalar 
Higgs bosons, and $H$ is the heavier (second lightest) neutral Higgs boson.
We assume that the Yukawa couplings of the other Higgs bosons 
are suppressed by the mixing angles \footnote{
This assumption is not compulsory, since all the Higgs bosons might 
participate in the top-quark pair production in principle. We will keep only
a few lightest (pseudo) scalar bosons in order to simplify the numerical analysis. 
}.


Introducing $U(1)'$ flavored Higgs doublets is very important because they generates 
nonzero top mass.  They also play an important role in top FBA phenomenology.  
For example the Yukawa couplings of the neutral scalar bosons $h,H,a$ have flavor 
changing  couplings to the up-type quarks because of the flavor non-universal nature 
of $Z'$ interaction~\cite{u1models}:
\begin{eqnarray} 
Y^h_{tu} & = & \frac{ 2m_t (g^u_R)_{ut} }{v \sin( 2 \beta)} 
\sin (\alpha-\beta) \cos \alpha_{\Phi} \ ,  
\\
Y^H_{tu} & = & - \frac{ 2m_t (g^u_R)_{ut} }{v \sin( 2 \beta)} 
\cos (\alpha-\beta) \cos \alpha_{\Phi} \ ,
\\
Y^a_{tu} & = & \frac{ 2m_t (g^u_R)_{ut} }{v \sin( 2 \beta)} \ .
\end{eqnarray} 
These Yukawa couplings are not present in the Type-II 2HDM, for example.  
Our models  proposed in Ref.~\cite{u1models} are good examples of non-minimal 
flavor violating multi-Higgs doublet  models, where the non-minimal flavor violation 
originates from the flavor non-universal chiral couplings of the new gauge 
boson $Z'$.   In our model,  the top FBA and the same-sign top-pair productions are 
generated not only by the $t$-channel $Z'$ exchange, but also by the $t$-channel 
exchange of neutral Higgs scalars,  and the strong constraint on the original $Z'$ 
model  from the same-sign top-pair production can be relaxed by a significant 
amount when we include all the contributions in the model, as described 
in the following section. 



\section{Phenomenology}

\subsection{Generalities and Inputs}

In this section, we discuss phenomenology of our model described in the previous 
section. If new physics affects the top-quark pair production and could accommodate  
$A_\textrm{FB}^t$ at the Tevatron,  it must also be consistent with many other 
experimental measurements related with the top quark.
In our models, both the $Z^\prime$ and Higgs bosons $h$ and $a$  contribute
to the top-quark pair production through the $t$-channel exchange in 
the $u\bar{u} \to t \bar{t}$ process.  As we discussed in the previous section,
the $Z^\prime$ boson also  contributes to the top-quark pair production 
through the $s$-channel exchange, which was ignored in Ref.~\cite{zprime}.

As two extreme cases, one can consider the cases where only the $Z^\prime$ boson
or Higgs boson $h$ contributes to the top-quark pair production.
Then, our models become close to the simple $Z^\prime$ model of Ref.~\cite{zprime}
or the scalar-exchange model of Ref.~\cite{babu}. 
Unfortunately, these models cannot be compatible with the present upper bound 
on the same-sign top-quark pair production at the LHC in the parameter space 
which give rise to a moderate $A_\textrm{FB}^t$~\cite{u1models}. 
In our chiral U(1)$^\prime$ models, the constraint from the same-sign top-quark
pair production could be relaxed because of the destructive interference
between the contribution from the $Z^\prime$ and those from Higgs bosons $h$ 
and $a$.  In particular, the contribution of the pseudoscalar boson $a$ to the 
same-sign  top-quark pair production is opposite to the other contributions.

In the two Higgs doublet model with the $U(1)^\prime$ assignments  
to the right-handed up-type quarks, $(u_i)=(0,0,1)$,  
the $s$-channel contribution of the $Z^\prime$ exchange 
to the partonic process  $u\bar{u}\to t\bar{t}$  is as strong as 
an $t$-channel contribution because of the relation 
$(g_R^u)_{ut}^2 = (g_R^u)_{uu} (g_R^u)_{tt}$~\cite{u1models}.
In the multi-Higgs doublet models (mHDMs)  with other U(1)$^\prime$ charge 
assignments $(u_i)'$s to the right-handed up-type quarks, 
the $s$-channel contribution could be small.
In general, one can write $(g_R^u)_{uu} (g_R^u)_{tt} = \xi (g_R^u)_{ut}^2$,
where $\xi$ is a function of mixing angles and $0\leq |\xi| \leq O(1)$. 
In the case of $m_{Z^\prime} \geq 2 m_t$, a resonance around the $Z^\prime$ 
mass for nonzero $\xi$ would be observed in the $t\bar{t}$ invariant mass 
distribution.  However, such a resonance has not been observed so far in the experiments.  This would restrict the $Z^\prime$ mass to be 
much smaller than $2 m_t$ for nonzero $\xi$.

The cross sections for the top-quark pair production at the Tevatron are 
$\sigma(t\bar{t})=(7.5\pm 0.48)$ pb at CDF~\cite{cdfttbar} and 
$\sigma(t\bar{t})= (7.56^{+0.63}_{-0.56})$ pb at D0~\cite{d0ttbar}, 
respectively. At the LHC, the cross sections for the top-quark pair production
are $\sigma(t\bar{t})=(165.8\pm 13.3)$ pb at CMS~\cite{cmsttbar} and 
$\sigma(t\bar{t})=(177\pm 11)$ pb at ATLAS~\cite{atlasttbar}, respectively.
In this work, we require that the cross section for the top-quark pair production
is in agreement with the CDF result in the $1\sigma$ level, which has
the least uncertainty. Another reason to use the Tevatron result for the check
of our model is that the top-quark pair production at the Tevatron is more
sensitive to new physics in the $u\bar{u}\to t\bar{t}$ process than at the LHC.

In the SM, the top quark dominantly decays into $W+b$. In our models, there are
several flavor-changing vertices $u_R$-$t_R$-$Z^\prime$, $u_R$-$t_L$-$h$, 
and $u_R$-$t_L$-$a$. If the $Z^\prime$ or Higgs bosons are lighter than the top 
quark, it could be dangerous because the branching ratio of the top quark
to $W+b$ is significantly altered. We assume that the pseudoscalar Higgs boson 
$a$ is heavier than the top quark and the branching ratio of 
the exotic decay of the top quark such as $t \rightarrow Z^\prime u , h u$ 
is less than 5\%. 
We find that  the exotic decay mode of the top quark can be suppressed, 
if we choose $\alpha_x \equiv ((g^\prime g_R^u)_{ut} )^2/(4\pi) \lesssim 0.012$ 
for $m_{Z^\prime} = 145$ GeV and $Y_{tu} \lesssim 0.5$ for $m_h=125$ GeV. 

Furthermore,  such large FCNCs could generate the same-sign top-quark pair 
production  through the $t$-channel diagram in the $uu \to tt$ process,  
which is forbidden within the SM. 
The CMS Collaboration announced the upper bound on the cross section
for the same-sign top-quark pair production: 
$\sigma^{tt} < 17$ pb at 95\% CL. with a luminosity of
35 pb$^{-1}$~\cite{cmssame},
while the limit on the cross sections at ATLAS with a luminosity of 
$1.04$ fb$^{-1}$ are
$\sigma^{tt} < 2$ pb at 95\% CL. by using an optimized event selection 
for the $Z^\prime$ model and
$\sigma^{tt} < 4$ pb at 95\% CL. by using more inclusive selection,
respectively~\cite{atlassame}. We use the latter limit in this work.

In numerical analysis, we take the top-quark mass to be $m_t=173$ GeV.
For a parton distribution function we use CTEQ6m with the renormalization
and the factorization scale equal to $\mu=m_t$. 
In order to take into account the QCD radiative correction which is unknown 
as of now for the model under consideration, we use the $K$ factor obtained 
in the perturbative QCD calculations: namely, $K=1.3$ for the Tevatron and 
$K=1.7$ for the LHC by assuming the same $K$ factor in the new physics model. 
The center-of-momentum energy $\sqrt{s}$ is $1.96$ TeV at the Tevatron and 
$7$ TeV at the LHC, respectively.
In the previous works~\cite{u1models}, we did not consider 
the SM NLO contribution to $A_\textrm{FB}^t$, but in this work
we take into account its contribution to $A_\textrm{FB}^t$ by using
the approximated formula $A_\textrm{FB}^t \simeq A_\textrm{FB}^{t,\textrm{SM}}
+\delta A_\textrm{FB}^{t}/K$, where the first term denotes
$A_\textrm{FB}^t$ at the SM NLO and the second one corresponds to
the contribution from the new physics. 
We also use the approximated formula $A_C^y \simeq A_C^{y,\textrm{SM}}
+\delta A_C^y/K$.

\subsection{$m_{Z'} = 145$ GeV and $\xi=1$}

In this model, the $Z^\prime$ boson can contribute to the top-quark pair
production through its $s$-channel and $t$-channel exchanges
in the $u\bar{u}\to t\bar{t}$ process. While the Higgs bosons contribute
to the top-quark pair production
only in the $t$ channel because the diagonal elements of their Yukawa couplings 
to light quarks   are negligible.
We scan the following parameter regions:
$180~\textrm{GeV} \le m_{H,a} \le 1~\textrm{TeV}$,
$0.005 \le \alpha_x \le 0.012$,
$0.5 \le Y_{tu}^{H,a} \le 1.5$, and
$(g_R^u)_{tu}^2=(g_R^u)_{uu} (g_R^u)_{tt}$, 
where $\alpha_x \equiv (g_R^u)_{tu}^2 g'^2/(4 \pi)$ is defined and $Y_{tu}^{H,a}$ are flavor-off-diagonal 
Yukawa couplings. 

In Fig.~1, we show the scattered plot for $A_\textrm{FB}^t$ at the Tevatron and 
the same-sign top-pair production cross section and $A_C^y$ 
at the LHC. 
The green and yellow regions are consistent with $A_C^y$ at ATLAS and CMS
in the $1\sigma$ level, respectively. The blue and skyblue regions are
consistent with $A_{FB}^t$ in the lepton+jets channel at CDF in the
$1\sigma$ and $2\sigma$ levels, respectively.
The red points are in agreement with the cross section 
for the top-quark pair production
at the Tevatron in the $1\sigma$ level and the blue points are consistent with
both the cross section for the top-quark pair production at the Tevatron
in the $1\sigma$ level and the upper bound on the same-sign top-quark pair
production at ATLAS. We find that a lot of parameter points can explain all
the experimental data.
We emphasize that the simple $Z^\prime$ model is excluded by the same-sign 
top-quark pair production, but in the chiral U(1)$^\prime$ model, this strong bound 
could be evaded  due to the destructive interference between the $Z^\prime$ boson 
and Higgs bosons.  Also the $m_{t\bar{t}}$ distribution becomes closer to the SM 
case in the presence of $h$ and $a$ contributions (see Fig.~2).  One can realize that it is important to include the Higgs contributions 
 as well as the $Z'$ contributions. All the physical observables are affected by the Higgs contributions.

\begin{figure}[!t]
\begin{center}
\epsfig{file=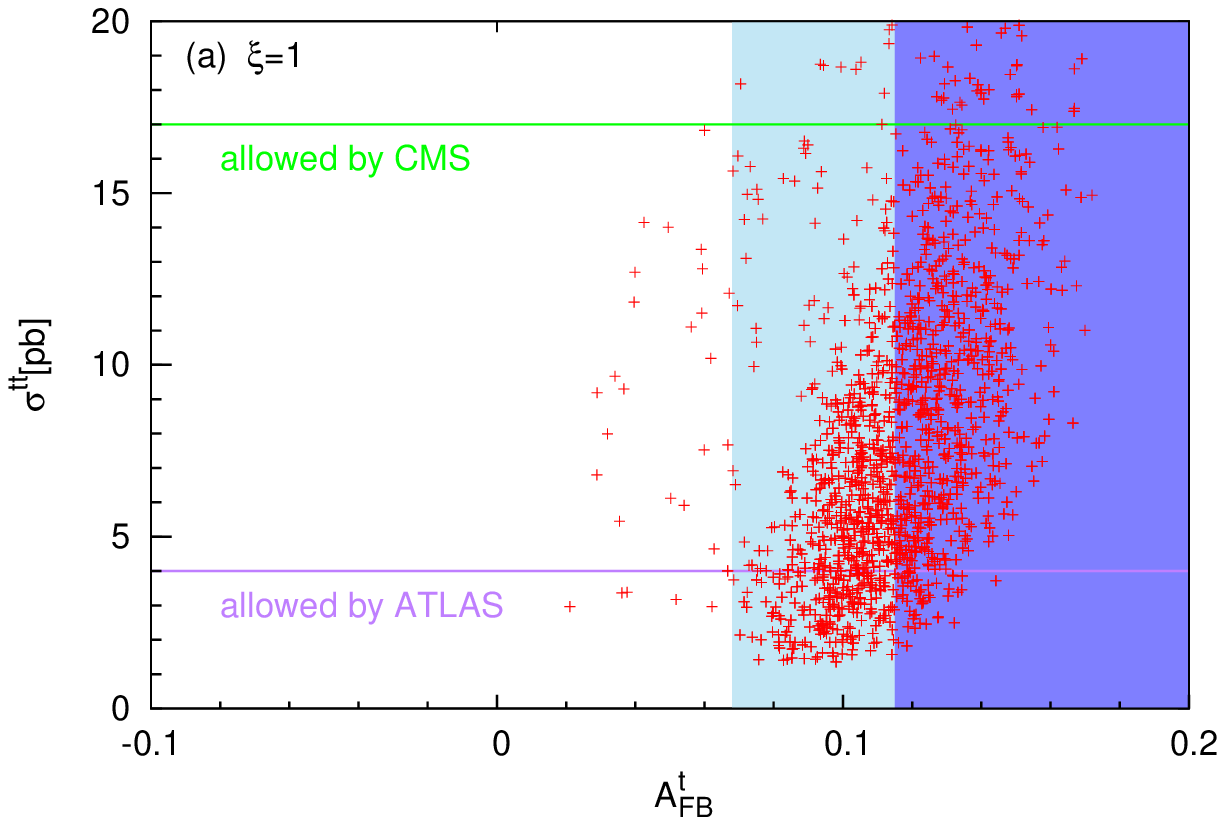,width=0.45\textwidth}
\epsfig{file=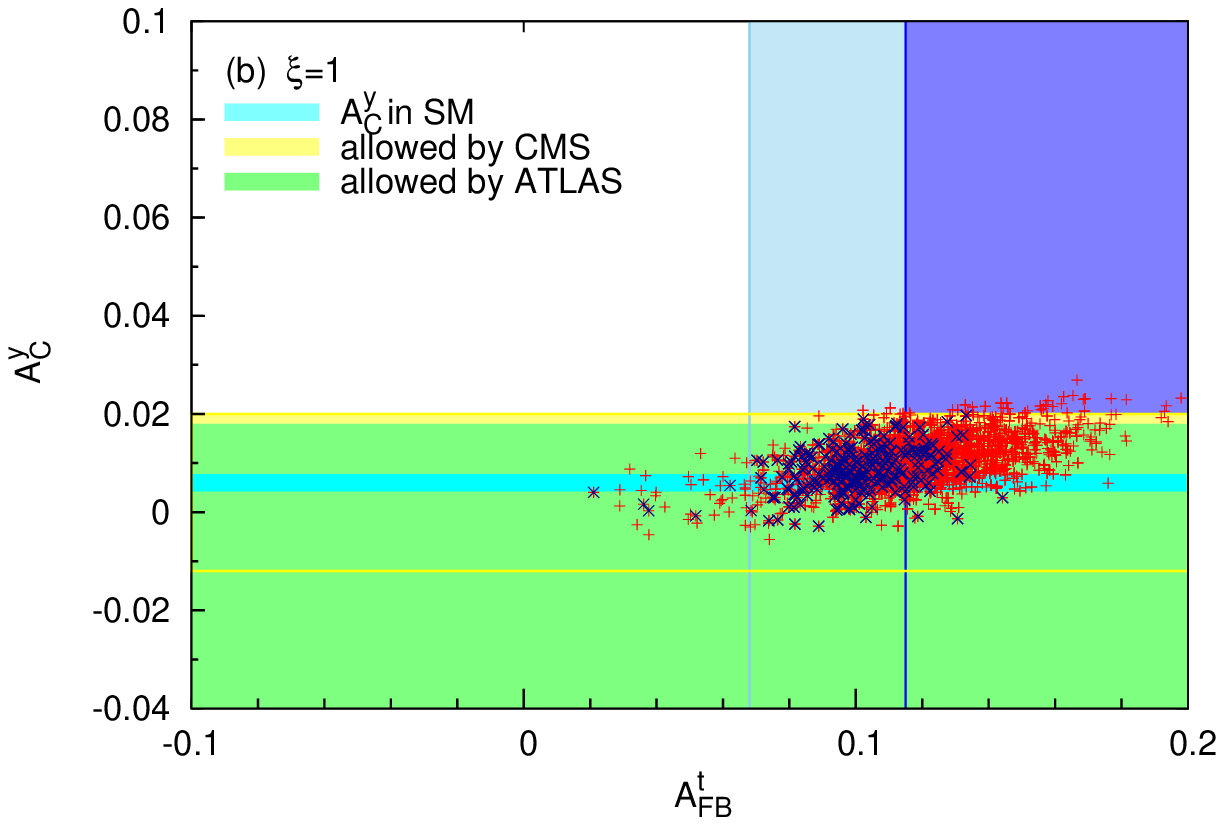,width=0.45\textwidth}
\caption{\label{fig:lightzp}%
The scattered plots for (a) $A_\textrm{FB}^t$ at the Tevatron and
$\sigma^{tt}$ at the LHC in unit of pb, and (b) $A_\textrm{FB}^t$ at the 
Tevatron and $A_C^y$ at the LHC for $m_{Z^\prime}=145$ GeV and $\xi=1$.
In (b), the blue points satisfy the upper bound on the same sign top pair 
production from ATLAS: $\sigma^{tt} < 4$ pb.
}
\end{center}
\end{figure}

\begin{figure}[!t]
\begin{center}
\epsfig{file=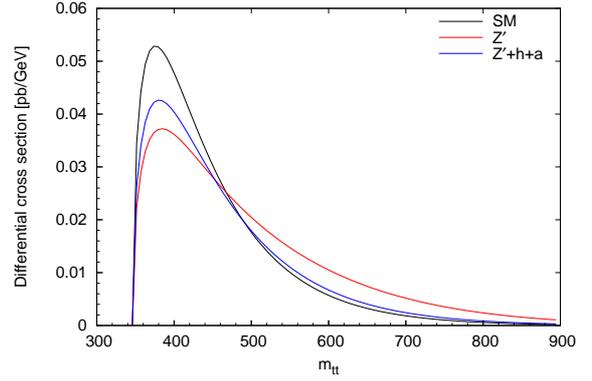,width=0.45\textwidth}
\caption{\label{fig:mtt}%
The invariant mass distribution of the top-quark pair at the Tevatron
in the SM, $Z^\prime$ model, and chiral U(1)$^\prime$ model.
}
\end{center}
\end{figure}

\subsection{$m_h = 125$ GeV and $\xi=0$}
In this subsection, we discuss the scenario that
 a light Higgs boson $h$ with $m_h=125$ GeV, motivated by the recent observation 
 of an SM-Higgs like scalar boson 
at the LHC~\cite{higgs}, also has a nonzero $Y^h_{tu}$. 
In this case, the $Z^\prime$ boson and Higgs bosons
$h$, $H$, and $a$ contribute to the top-quark pair production.
In order to suppress the exotic decay of the top quark into $h$ and $u$, 
we set the Yukawa
coupling of $h$ to be $Y_{tu}^h \le 0.5$ and masses of $Z^\prime$, $H$,
and $a$ are larger than the top-quark mass or approximately equal to the
top-quark mass. We scan the following parameter regions:
$160~\textrm{GeV} \le m_{Z^\prime} \le 300~\textrm{GeV}$,
$180~\textrm{GeV} \le m_{H,a} \le 1~\textrm{TeV}$,
$0 \le \alpha_x \le 0.025$,
$0 \le Y_{tu}^{H,a} \le 1.5$, $0 \le Y_{tu}^{h} \le 0.5$ and $\xi=0$. 
The mass region of the $Z^\prime$ boson
is taken to avoid the constraint from the $t\bar{t}$ invariant mass
distribution at the LHC. If $(g_R^u)_{uu}\simeq0$ 
and the $s$-channel contribution of the $Z^\prime$ could be ignored,
the mass region of the $Z^\prime$ boson could be enlarged.
In Fig.~3, we show the scattered plot for $A_\textrm{FB}^t$ 
at the Tevatron and $A_C^y$ at the LHC for $m_h=125$ GeV.
All the legends on the figure are the same as those in Fig.~1.
We find that there exist parameter regions which agree with all the experimental
constraints. 
We emphasize that in some parameter spaces $\sigma^{tt}$ is less than 1 pb.

\begin{figure}[!th]
\begin{center}
\epsfig{file=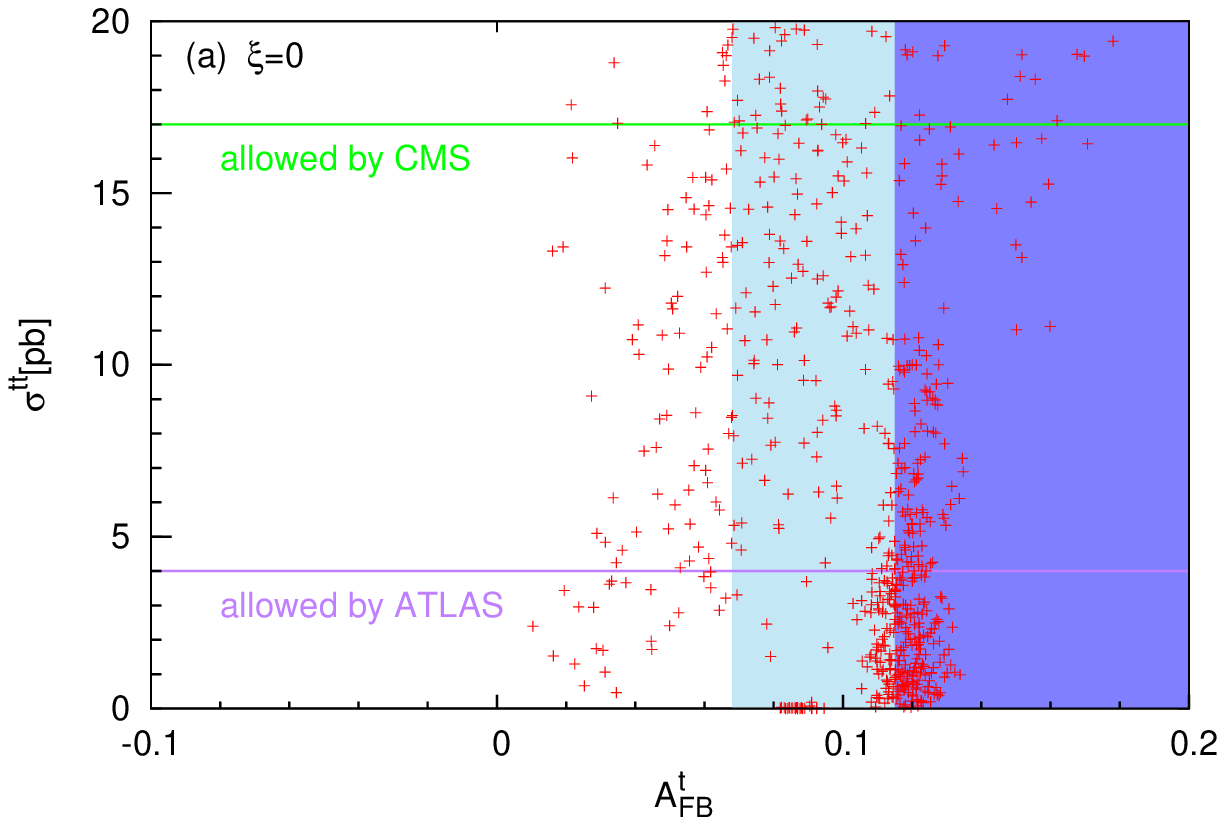,width=0.45\textwidth}
\epsfig{file=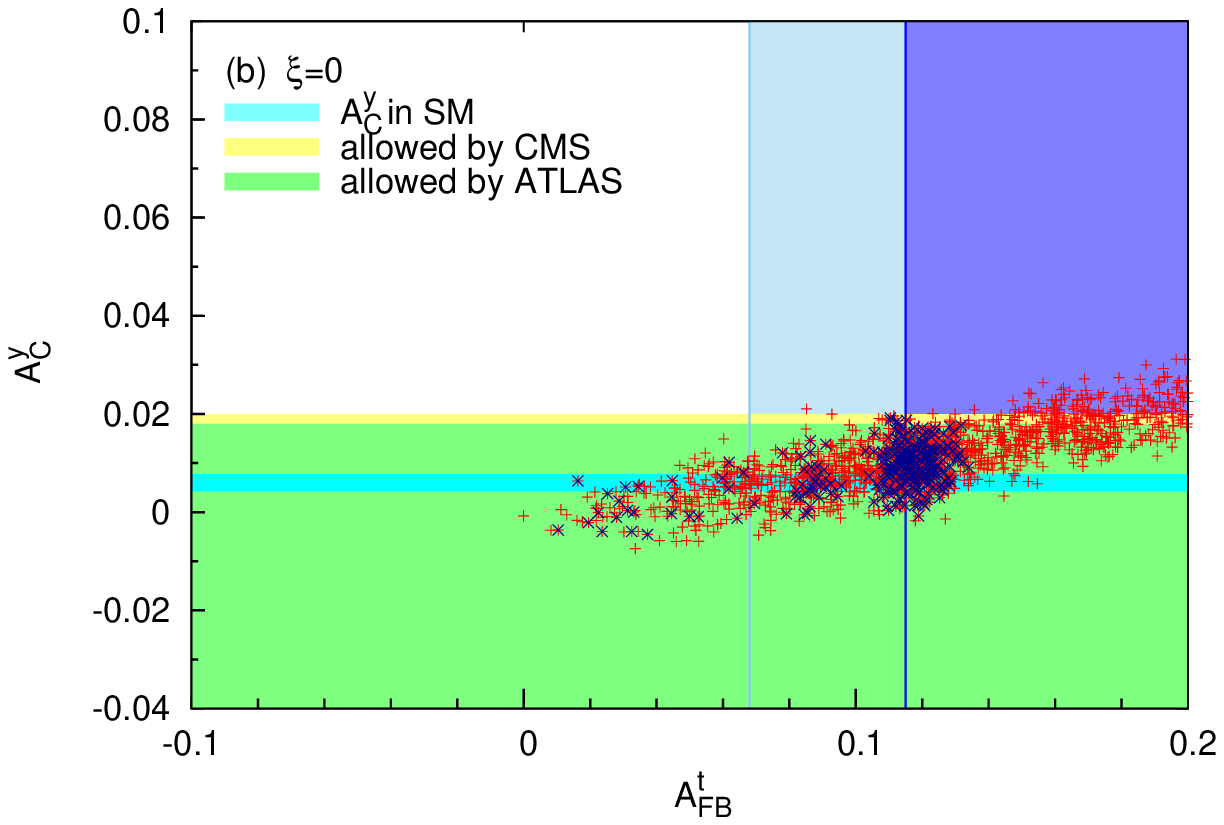,width=0.45\textwidth}
\caption{\label{fig:higgs2-2}%
The scattered plots for (a) $A_\textrm{FB}^t$ at the Tevatron and
$\sigma^{tt}$ at the LHC in unit of pb, and (b) $A_\textrm{FB}^t$ at the 
Tevatron and $A_C^y$ at the LHC for $m_h=125$ GeV and $\xi=0$,
where the contribution of the second lightest Higgs boson $H$ is included.
}
\end{center}
\end{figure}

\section{Summary}
The top forward-backward asymmetry at the Tevatron
is the only quantity which has 
deviation from the SM prediction in the top quark sector up to now.
A lot of new physics models have been introduced to 
account for this deviation.  
Or it has been analyzed in a model-independent way~\cite{modelindep,grojean}, 
and some models have already been disfavored by experiments at the LHC.
In this work, we investigated the chiral U(1)$^\prime$ model with flavored
Higgs doublets and flavor-dependent couplings. Among possible scenarios,
we focused on two scenarios in Sec.~4. 
We found that both scenarios can be accommodated with
the constraints from the same-sign top-quark pair production and
the charge asymmetry at the LHC as well as the top forward-backward asymmetry
at the Tevatron.

The chiral U(1)$^\prime$ model has a lot of new particles except for
the $Z^\prime$ boson and neutral Higgs bosons. The search for exotic particles
may constrain our model severely. For example, our model is strongly
constrained by search for the charged Higgs boson in the
$b\to s\gamma$, $B\to \tau \nu$, and $B\to D^{(\ast)}\tau \nu$ decays \cite{bdecay}.
In order to escape from such constraints, we must assume a quite heavy charged
Higgs boson or it is necessary to study our model more carefully
by including all the interactions which have been neglected in this work.
More detailed analysis on this issue can be found in Ref.~\cite{bdecay}.

\section*{Acknowledgments}
This work is supported in part by Basic Science Research Program
through NRF 2011-0022996 and in part by NRF Research Grant
2012R1A2A1A01006053.



\begin{thebibliography}{}   

\bibitem{cdfnew}
  CDF Collaboration,
  CDF Conf. note {\bf 10807} (2012).

\bibitem{oldafb}
  T.~Aaltonen {\it et al.}  [CDF Collaboration],
  Phys.\ Rev.\  D {\bf 83}, 112003 (2011);
  CDF Collaboration,
  CDF note {\bf 10436} (2011);
  V.~M.~Abazov {\it et al.}  [D0 Collaboration],
  Phys.\ Rev.\ D {\bf 84}, 112005 (2011).  


\bibitem{smafbac}
  O.~Antunano, J.~H.~Kuhn and G.~Rodrigo,
  Phys.\ Rev.\  D {\bf 77}, 014003 (2008).

\bibitem{smafb}
  V.~Ahrens, A.~Ferroglia, M.~Neubert, B.~D.~Pecjak and L.~L.~Yang,
  Phys.\ Rev.\  D {\bf 84}, 074004 (2011);
  W.~Hollik and D.~Pagani,
  Phys.\ Rev.\  D {\bf 84}, 093003 (2011);
  J.~H.~Kuhn and G.~Rodrigo, JHEP {\bf 1201}, 063 (2012).  


\bibitem{atlasacy}
  ATLAS Collaboration,
  ATLAS-CONF-2011-106 (2011).

\bibitem{cmsacy}
  CMS Collaboration, CMS-PAS-TOP-11-030 (2011).

\bibitem{cmssame}
  S.~Chatrchyan {\it et al.}  [CMS Collaboration], JHEP {\bf 1108}, 005 (2011).


\bibitem{atlassame}
  G.~Aad {\it et al.}  [ATLAS Collaboration], JHEP {\bf 1204}, 069 (2012).


\bibitem{Chatrchyan:2012sa} 
S.~Chatrchyan {\it et al.}  [CMS Collaboration],
  JHEP {\bf 1208}, 110 (2012)
  [arXiv:1205.3933 [hep-ex]].


\bibitem{zprime}
  S.~Jung, H.~Murayama, A.~Pierce and J.~D.~Wells,
   Phys.\ Rev.\  D {\bf 81}, 015004 (2010)
  [arXiv:0907.4112 [hep-ph]].

\bibitem{saavedra}
J.~A.~Aguilar-Saavedra and M.~Perez-Victoria,
  Phys.\ Lett.\ B {\bf 701}, 93 (2011)
  [arXiv:1104.1385 [hep-ph]]; 
  JHEP {\bf 1109}, 097 (2011)
  [arXiv:1107.0841 [hep-ph]].

\bibitem{u1models}
  P.~Ko, Y.~Omura and C.~Yu,
  Phys.\ Rev.\ D {\bf 85}, 115010 (2012);  
  JHEP {\bf 1201}, 147 (2012);
  Nuovo Cim.\ C {\bf 035N3}, 245 (2012); 
  arXiv:1205.0407 [hep-ph].  

\bibitem{babu}
 K.~S.~Babu, M.~Frank and S.~K.~Rai,
  Phys.\ Rev.\ Lett.\  {\bf 107}, 061802 (2011)
  [arXiv:1104.4782 [hep-ph]].

\bibitem{cdfttbar}
  CDF Collaboration, CDF note {\bf 9913} (2009).

\bibitem{d0ttbar}
  V.~M.~Abazov {\it et al.}  [D0 Collaboration],
  Phys.\ Lett.\  B {\bf 704}, 403 (2011)
  [arXiv:1105.5384 [hep-ex]].

\bibitem{cmsttbar}
  CMS Collaboration, CMS-PAS-TOP-11-024 (2011).

\bibitem{atlasttbar}
  ATLAS Collaboration, ATLAS-CONF-2012-024 (2012).

  
\bibitem{higgs}
  G.~Aad {\it et al.}  [ATLAS Collaboration],
  arXiv:1207.7214 [hep-ex];  
 J.~Incandela,
 CMS talk at {\it Latest update in the search for the Higgs boson} at CERN,
 July 4, 2012.



\bibitem{modelindep}
  D.~W.~Jung, P.~Ko, J.~S.~Lee and S.~h.~Nam,
  Phys.\ Lett.\  B {\bf 691}, 238 (2010);
  D.~W.~Jung, P.~Ko and J.~S.~Lee,
   Phys.\ Lett.\  B {\bf 701}, 248 (2011);
  Phys.\ Rev.\  D {\bf 84}, 055027 (2011);
  Phys.\ Lett.\  B {\bf 708}, 157 (2012).
  
 \bibitem{grojean}
  C.~Degrande, J.~-M.~Gerard, C.~Grojean, F.~Maltoni and G.~Servant,
  JHEP {\bf 1103}, 125 (2011)
  [arXiv:1010.6304 [hep-ph]].
  
  \bibitem{bdecay}
P.~Ko, Y.~Omura and C.~Yu,
  arXiv:1212.4607 [hep-ph].
\end{thebibliography}
  

%
%

\end{document}